\documentstyle[12pt,aaspp4]{article}

\newcommand{\kms}{km~s$^{-1}$}
\newcommand{\tco}{$^{13}$CO}
\newcommand{\um}{$\mu$m}

\received{2003 October 1}
\begin{document}
\accepted{  }
\journalid{ }{ }
\slugcomment{   }
\lefthead{Plume et al.}
\righthead{Water Absorption From Line-of-Sight Clouds Toward W49A}

\title{Water Absorption From Line-of-Sight Clouds Toward W49A}

\author{
Ren\'e Plume\altaffilmark{1},
Michael J. Kaufman\altaffilmark{2},
David A. Neufeld\altaffilmark{3},
Ronald L. Snell\altaffilmark{4},
David J. Hollenbach\altaffilmark{5},
Paul F. Goldsmith\altaffilmark{6},
John Howe\altaffilmark{4},
Edwin A. Bergin\altaffilmark{7}, 
Gary J. Melnick\altaffilmark{7}, 
and
Frank Bensch\altaffilmark{7}
} 

\altaffiltext{1}{Department of Physics \& Astronomy, University of Calgary, 2500 University Dr. NW,
Calgary, AB, T2N-1N4, Canada}
\altaffiltext{2}{Department of Physics, San Jose State University, One Washington Square, San Jose, CA 95192}
\altaffiltext{3}{Department of Physics and Astronomy, Johns Hopkins
University, 3400 N. Charles Street, Baltimore MD 21218}
\altaffiltext{4}{Department of Astronomy,  University of Massachussetts, Amherst, Lederle Graduate Research Tower, MA 01003}
\altaffiltext{5}{NASA Ames Research Center, MS 245-3, Moffett Field, CA 94035}
\altaffiltext{6}{NAIC, Department of Astronomy, Cornell University, Ithaca, NY 14853}
\altaffiltext{7}{Harvard--Smithsonian Center for Astrophysics,
       60 Garden St., Cambridge, MA 02138}


\begin{abstract}

We have observed 6 clouds along the line-of-sight toward W49A using the Submillimeter
Wave Astronomy Satellite (SWAS) and several ground-based observatories.  The ortho-H$_2$O $1_{10} \rightarrow 1_{01}$ 
and OH (1665 and 1667 MHz) transitions are observed in {\it absorption}, whereas the low-J
CO, $^{13}$CO, and C$^{18}$O lines, as well as the [C~I] $^3P_1 - ^3P_0$ transition, are seen in {\it emission}.  
The emission lines allow us to determine the gas density (n $ \sim  1500 - 3000$ cm$^{-3}$) and CO
column densities ($N(CO) \sim 7.9 \times 10^{15} - 2.8 \times 10^{17}$ cm$^{-2}$) using a standard Large Velocity Gradient analysis.  
 
By using both the o-H$_2^{18}$O and o-H$_2$O absorption lines, we are able to constrain the column-averaged o-H$_2$O 
abundances  in each line-of-sight cloud to within about an order of magnitude.  
Assuming the standard N(H$_2$)/N(CO) ratio of $10^4$, we find  N(o-H$_2$O)/N(H$_2$) = $8.1 \times 10^{-8} - 4 \times
10^{-7}$ for three clouds with optically thin water lines.  In three additional clouds, the H$_2$O lines are saturated so we have used observations of the H$_2^{18}$O ground-state transition to find  upper limits to the water abundance of $8.2\times 10^{-8} - 1.5\times10^{-6}$.
We measure the OH abundance  from the average of the 1665 and 1667 MHz observations and find   N(OH)/N(H$_2$) = $2.3 \times 10^{-7} - 1.1 \times 10^{-6}$.
The o-H$_2$O and OH abundances are similar to those determined for line-of-sight water absorption 
features towards W51 and Sgr B2  but are higher than those seen from water
 {\it emission} lines in molecular clouds.  However, the clouds towards W49 have lower ratios of OH
relative to H$_2$O column densities than are predicted by simple models which assume that dissociative recombination is the primary formation pathway for OH and H$_2$O.  Building on the work of Neufeld et al. (2002),
we present photo-chemistry models including additional chemical effects, which can also explain the observed OH and H$_2$O column densities as well as the observed H$_2$O/CO abundance ratios.

\end{abstract}


\section{Introduction}

Since its launch in 1998 December, the Submillimeter Wave
Astronomy Satellite (SWAS) has observed emission from the 
ground-state ($1_{10} \rightarrow 1_{01}$) transition of ortho-H$_2$O (hereafter o-H$_2$O) in
a number of molecular clouds (e.g. Melnick et al. 2000a; Snell et al.
2000a, 2000b). While it has been possible to 
determine the H$_2$O abundances from the SWAS observations, these 
analyses depend on the often poorly contrained gas densities and
temperatures.  Observations of {\it absorption} from the o-H$_2$O ground-state 
transition, however, have the advantage that the
column density is simply proportional to the optical
depth in the line. Therefore, the H$_2$O abundance can be
derived without precise knowledge of
the temperature or density of the absorbing gas.

In this paper we present observations of water absorption toward 
the molecular cloud complex W49A.  
W49 is 11.4 kpc from the Sun and 
8.1 kpc from the Galactic Center (Gwinn et al. 
1992).  It consists of a supernova remnant (W49B) and an HII region (W49A)
separated by $\sim 12'$.  W49A has long been known as a site of 
extremely active star formation,
due to its association with one of the most powerful H$_2$O masers in the 
Galaxy (Genzel et al. 1978), and the fact that it is 
one of the most luminous regions in the Galaxy 
($L_{bol} \sim 10^7$ L$_\odot$; Ward-Thompson \& Robson 1990).
Physically, W49A is composed of a number of optically obscured, 
compact HII regions
surrounded by a massive molecular cloud ($M > 10^5$ M$_{\odot}$; Mufson \&
Liszt 1977).
W49A is broken up into three main IR peaks; W49SW, W49SE and, the strongest,
W49NW (more commonly named W49N; Harvey et al. 1977).  The continuum emission from
W49A has also been observed at 350, 800, and 1100 $\mu$m by Ward-Thompson
and Robson (1990) and at 1300 $\mu$m by Schloerb et al. (1987).

Spectrally, W49A is extremely complex,
containing numerous features contributed by W49A itself, as well as
additional clouds associated 
with the Sagittarius spiral arm (which crosses
the line-of-sight twice).  W49A and the associated line-of-sight clouds
have been the subject of numerous spectroscopic studies (both emission
and absorption) involving species such as HI (Radhakrishnan et al. 1972), HCO$^+$ and
HCN (Nyman 1983), OH (Cohen and Willson 1981), H$_2$CO (Mufson and Liszt 1977;
Bieging et al. 1982), CO (Mufson and Liszt 1977), SiO (Lucas and Liszt 2000),
CS (Greaves and Williams 1994), and OI (Vastel et al. 2000).

The strong continuum provided by this source, and the well-studied 
deep absorption features in the gas associated with 
line-of-sight clouds, provide an excellent opportunity to 
measure the H$_2$O abundances in a number of different molecular clouds. 
Similar analyses of the water abundances have recently been carried out for line-of sight
clouds toward  Sagittarius B2
( Neufeld et al. 2000; Cernicharo et al. 1997), Sagittarius A* (Moneti et al. 2001), and W51 (Neufeld et al. 2002).

\section{Observations}

Since SWAS has two independent receivers, we were able to
simultaneously observe the 
[C~I] $^3P_1 \rightarrow ^3P_0$ ($\nu = 492.1607$ GHz) and O$_2$ $3,1 \rightarrow 3,2$ 
($\nu = 487.249$ GHz) transitions
 in the upper and lower sidebands of receiver 1, and the 
$^{13}$CO ~$J = 5\rightarrow 4$    ($\nu = 550.926$ GHz) and 
o-H$_2$O $1_{10}\rightarrow 1_{01}$ ($\nu = 556.936$ GHz) 
transitions in the upper and lower sidebands of receiver 2.
However, when receiver 2 was tuned to the 
o-H$_2^{18}$O $1_{10}\rightarrow 1_{01}$ transition ($\nu = 547.676$ GHz), SWAS loses
the ability to observe o-H$_2$O and $^{13}$CO $J = 5\rightarrow 4$.
The integration times and $1\sigma$ rms noise levels for each line are
listed in Table 1.
The SWAS beamsize at 557 GHz is $3.3'\times 4.5'$ while at 490 GHz it
is $3.5'\times 5.0'$.  The 
main beam efficiency ($\eta_{mb}$) for SWAS is 90\%.  
For more information about the SWAS instrument, data acquisition,
and data analysis, see Melnick et al. (2000b).

The SWAS observations of W49 were obtained in four different observing
periods; 1999 April and May, 1999 September and October, 
2000 March - May, and 2000 October.
The central position (located at W49N) is given as 
$\alpha(J2000) = 19^h 10^m 13.5^s$, $\delta(J2000) = 09^{\circ}:06':29''$.
The reference position  was $\sim 1.5^{\circ}$ from the central position
($\alpha(J2000) = 19^h 15^m 19.8^s$, $\delta(J2000) = 08^{\circ}:15':22''$) and
was chosen based on an absence of $^{12}$CO ~$J = 1 \rightarrow 0$ emission.

In order to understand the H$_2$O and OH absorption, it is important to know the spatial extent of the
submillimeter continuum.  To this end, we have mapped a region in W49 comparable to the SWAS beam at
450 $\mu$m using the SHARC camera at the Caltech Submillimeter
Observatory (CSO).  These observations were
taken in 1999 September.  The flux levels were calibrated using Uranus
as the calibration source. In addition to the SHARC maps, we
also have data from the on-board SWAS continuum
detectors in the fast-chop (4 Hz) observing mode (see Melnick et al. 2000b
for details).

We observed $^{12}$CO $J = 1\rightarrow 0$ ($\nu = 115.271$ GHz),  $^{13}$CO  $J = 1\rightarrow 0$ 
($\nu = 110.201$ GHz), and C$^{18}$O ~$J = 1\rightarrow 0$ ($\nu = 109.782$ GHz),
at the 14m Five Colleges Radio Astronomy Observatory (FCRAO) using their 32 element array 
receiver (SEQUOIA).  The $^{12}$CO ~$J = 1\rightarrow 0$ observations were taken in 2001 August 
on an 8 beam $\times$ 8 beam grid with 44$''$ spacings (the resolution of the FCRAO antenna).  The  
$^{13}$CO and C$^{18}$O ~$J = 1\rightarrow 0$  observations were obtained in 2002 November.  
For comparison with the SWAS H$_2$O observations these data were smoothed to a resolution of 
1.5$'$ to match the size of the continuum source in W49 (see Section 3.1).  
 The main beam efficiency 
($\eta_{mb}$) of the FCRAO is 49\%

Observations of the $^{12}$CO ~$J = 3 \rightarrow 2$ transition were made at
the KOSMA (K\"olner Observatorium f\"ur Submm-Astronomie) 3m telescope near Zermatt,
Switzerland.  The observations were taken on 2000 November 19 using the MRS (medium 
resolution acousto optical spectrometer) backend with a resolution of 360 kHz, corresponding to 0.31 
km s$^{-1}$ at 345 GHz.  The beamsize at 345 GHz is 82$''$ and the main beam efficiency 
($\eta_{mb}$) is 78\%.  For comparison with the SWAS observations, these data were smoothed 
to match the spatial resolution of the SWAS $^{13}$CO ~$J = 5 \rightarrow 4$ observations.

In order to further examine the line-of-sight clouds towards W49A,  observations of the 
F = $1 \rightarrow 1$ (1665 MHz) and F = $2 \rightarrow 2$ (1667 MHz) transitions of OH 
were obtained at the Arecibo Observatory in 2001 December.  At these frequencies, the Arecibo 
beam size is approximately $2.6' \times 3'$ (FWHM) and has a main beam efficiency  at 1665 \& 
1667 MHz of 80\%.  We also obtained observations of the 1612 and 1720 MHz transitions
of OH at the Arecibo observatory in 2003 October allowing us to investigate
 LTE versus non-LTE excitation in the OH lines, 

The on-source integration times and 1 $\sigma$ rms noise levels for each of
the observed spectral lines are presented in Table 1.
All spectra in this paper are presented in units
of T$_A^*$.  For subsequent analysis we use the main beam temperature (T$_{mb}$) which is defined by the
standard equation  T$_{mb} = T_A^*/ \eta_{mb}$ where $\eta_{mb}$ is the 
main beam efficiency.

\section{Results}

Figure \ref{fig:cont} shows the 450 $\mu m $ continuum emission obtained using SHARC at the CSO.  
The continuum emission is strongly concentrated on W49A itself but with weaker tendrils of emission 
extended approximately $0.75'$ from the center.  We use the 450  $\mu$m map of W49 to 
confirm the
continuum level in the SWAS o-H$_2$O and o-H$_2^{18}$O spectra.  
First, we extrapolate the integrated 450 $\mu$m continuum emission 
to 540 $\mu$m (the wavelength of the SWAS o-H$_2$O line) assuming a power 
law of the form 
$S \propto \lambda^{-3.5}$.
Then, we convert the integrated flux at 540 $\mu$m to an
equivalent main beam brightness temperature (T$_{mb}$) through the
standard equation: $S = {{2kT_{mb}\Omega}\over{\lambda^2}}$. Using
parameters appropriate for the SWAS beam at 540 $\mu$m, we obtain
$T_{mb} \sim S(Jy)/13982$.  The integrated fluxes for W49 are 1827 Jy at 450 $\mu$m and  
965 Jy at 540 $\mu$m, corresponding to T$_{mb}$ = 0.07 K at 540 $\mu$m.

At  539 $\mu$m, the SWAS
on-board continuum detectors 
yield a T$_{mb}$ of 0.1 K for W49A.
The fact that the SWAS results are
$\sim$ 30\% higher than SHARC observations
probably reflects uncertainties in the extrapolation to 540 $\mu$m,  atmospheric effects, and the fact that
the SHARC maps do not perfectly match the SWAS beam size, shape,
and orientation.  It should be noted, however, that the SWAS measurements may not be entirely 
accurate, due to a relatively slow chop rate.
Therefore, for the subsequent analysis in this paper we instead use the fact that 
four of the H$_2$O absorption
features are at the same depth to within the noise (the $\sim 18$ km s$^{-1}$ component from W49 itself, and 
the 39.5, 59.6, and 63.3 km s$^{-1}$ features; see Figure   \ref{fig:spec1}).  This
strongly suggests that these lines are saturated and thus, the continuum flux level can be 
set by the depth of these four features.  The average depth of these four lines is 0.08 K ($T_A^*$) which agrees satisfactorily  with both our measurement from SWAS' on-board continuum detectors and our extrapolation of the SHARC data.

Figure \ref{fig:spec1} shows the o-H$_2$O $1_{10} \rightarrow 1_{01}$, [C~I]
$^3P_1 \rightarrow ^3P_0$, and $^{13}$CO ~$J = 5 \rightarrow 4$ observations from SWAS. 
Velocities less than  $\sim 30$ km s$^{-1}$ correspond to emission/absorption
intrinsic to W49A and these features are not further analysed in this paper.  At 
velocities greater than $\sim 30$ km s$^{-1}$, Figure \ref{fig:spec1} shows
six distinct o-H$_2$O absorption dips at LSR velocities of 33.5, 39.5, 53.5, 59.6, 
63.3 and 68 km s$^{-1}$.
Gaussian fits to the baseline subtracted spectra are listed in Table 2  for each of the water 
absorption features.  The same six features
are seen in emission in the [C~I] spectrum  (Figure \ref{fig:spec1}), although at slightly different
LSR velocities.  The slight offset between the [C~I] and o-H$_2$O line centers may be partly 
due to the high optical depths of the water lines.  The Gaussian fits to the 
baseline subtracted carbon lines are given in Table 3.
No $^{13}$CO $J = 5 \rightarrow 4$ lines were detected at velocities
$> 30$ \kms .  The O$_2$ transition was not detected to a $3\sigma$ level of 15 mK.

Figure \ref{fig:spec2} shows the  $^{12}$CO, $^{13}$CO and C$^{18}$O ~$J = 1 \rightarrow 0$ 
observations from FCRAO and the $^{12}$CO ~$J = 3 \rightarrow 2$ observations from
KOSMA.  All spectra have been convolved to a circular, Gaussian beam of $ 3.8'$ to match the beam resolution of the SWAS $^{13}$CO
$J = 5 \rightarrow 4$ observations .  Figure \ref{fig:spec2} shows that while all six H$_2$O
absorption features are seen in both CO lines and the $^{13}$CO ~$J = 1 \rightarrow 0$  lines, the 
less abundant C$^{18}$O ~$J = 1 \rightarrow 0$  only shows the 3
strongest components.  
It should be noted that, while the spectra in Figure \ref{fig:spec2} have been smoothed to match 
the spatial resolution of the SWAS beam,
for our subsequent  LVG analysis (Section 3.1), we use data smoothed to the size of the background continuum source (1.5$'$). Since 
we are trying to estimate the CO column density in the gas which is absorbing the H$_2$O, it is more 
relevant to coadd only those $^{13}$CO and C$^{18}$O spectra that fall in front of the continuum source.  
Therefore, we  coadd the $^{13}$CO and C$^{18}$O spectra within $\sim 0.75'$ of the central position 
(Figure \ref{fig:cont}).  Gaussian fits to the detected spectral
features in the data smoothed to 1.5$'$ (with first order baselines subtracted) are given in Table 4. 

Figure \ref{fig:linecont} plots the line to continuum flux ratio
for the SWAS observations of both o-H$_2$O and o-H$_2^{18}$O.
The line to continuum flux ratio is given by 
$F_l/F_c = [\Delta T_{mb}(line) + T_{mb}(cont)]/T_{mb}(cont)$ where $T_{mb}(cont)$ is
the measured  main beam, SSB continuum flux level 
(see above) and $\Delta T_{mb}(line)$ is the
 baseline subtracted  antenna temperature 
($\Delta T_A^*$; listed in Table 2) divided by the main beam efficiency.
Although there are no clear o-H$_2^{18}$O lines at velocities
$> 30$ \kms , we provide upper limits  in Table 2 by fitting Gaussians with
 $LSR$ velocities and $FWHM$ linewidths fixed at the values
provided by the  
[C~I] emission  lines. We used the [C~I] spectrum as a template, rather than the o-H$_2$O
spectrum, due to the large optical depths of the water line.  Therefore,
Table 2 provides limits to the o-H$_2^{18}$O line strengths
and integrated intensities.   Table 2 also lists the line to continuum flux ratios (seen in 
Figure  \ref{fig:linecont}) and the line center optical depth ($\tau_o = -ln(F_l/F_c)$).

Since the continuum level is set by the average depth of the four deepest absorption features,  the 39.5, 59.6, and 63.3 km s$^{-1}$ o-H$_2$O features in Figure \ref{fig:linecont} 
reach line-to-continuum values close to, or less than, zero (and are, in fact,  as strong as the absorption feature associated
with W49A itself).  Therefore, these features are most likely saturated, and we would consequently derive a null or negative line flux and an unphysical opacity.  For these three lines, we provide a {\it minimum} line center opacity in Table 2 by adding the $1 \sigma$ error of the Gaussian fits ( $T_A^*$ = 0.006 K)  to the measured continuum level.
The other three features (at 33.5,
53.5, and 68 km s$^{-1}$) are weaker and appear optically thin.  It is important to note however, that 
while we assume these components to be optically thin they may, in fact, have a higher optical depth.  
The relative line strengths of the six $^{13}$CO  ~$J = 1 \rightarrow 0$ components  show significant 
variation over the region mapped, implying the presence of cloud structure on scales smaller than the SWAS beam.  Therefore, we cannot rule out the possibility that the 33.5,
53.5, and 68 km s$^{-1}$ features are actually saturated and only appear to be optically thin because 
they do not completely cover the continuum source.  Nevertheless, for the subsequent 
analysis in this paper we will assume that these three components are unsaturated.

Figure \ref{fig:ohlinecont} plots the line to continuum flux ratio for the Arecibo observations 
of the OH 1665 MHz and 1667 MHz transitions.  Since the  front end amplifiers at the Arecibo 
observatory have uniquely defined pass bands, there is no single sideband to double sideband 
conversion issue.  Thus, the measured SSB continuum levels
($T_{mb}(cont)$)  are 311.3 K at 
1665 MHz and 315.3 K at 1667 MHz.  
All six of the features seen in emission in 
[CI] are seen in absorption in both OH lines.    Table 5 lists the Gaussian fit parameters to all 
six components (with a first order baseline removed), as well as the line to continuum flux 
ratios and associated line center optical 
depth.  The lines at 39.3, 54.2, 62.9, and 68.8 \kms\ are all within 25\% of the 5/9 integrated 
intensity ratio expected for the 1665/1667 lines in local thermodynamic equilibrium (LTE).  
The 1665/1667 ratios in the 33.6 and 59.8 \kms\ lines, however, are $\sim 1/3$ which is 
$\sim 40$\% lower than expected for LTE.  The weakness of these lines suggests that they are optically thin.  Therefore,
the 1665/1667 ratio of 1/3 seen in the 33.6 and 59.8 \kms\ lines is probably a result of non-LTE excitation and not a line saturation effect.
In addition, our most recent OH observations (at 1612, 1665, 1667, and 1720 MHz) taken at the Arecibo observatory strongly suggest that the 39.3 and 62.9 \kms\ features are also affected by non-LTE excitation.  Therefore, we will restrict further analysis of the OH absorption lines to the 54.2 and 68.8 \kms\ 
components.  

\subsection{CO and C$^o$ Abundances}

To determine the line-of-sight average H$_2$O and OH abundances, we first need to 
estimate the  CO column density corresponding to each of the six water absorption features.  To do this,
 we use a 
Large Velocity Gradient (LVG) code to simultaneously fit the  $^{13}$CO ~$J = 1 \rightarrow 0$,
 $5 \rightarrow 4$, and C$^{18}$O ~J = $1 \rightarrow 0$
observations.  Since $^{13}$CO ~$J = 5 \rightarrow 4$ emission was not detected in any of the clouds, 
we use the 1 $\sigma$ noise level in the spectrum as an upper limit to the line strength in order to better 
constrain the fitting routine.  We do not include the $^{12}$CO observations in this analysis due to the 
higher opacity in these transitions and chose to focus instead on the optically thin tracers.  An 
investigation into the effect of ignoring the $^{12}$CO data in the LVG fits shows that the derived 
densities and column densities change by less than a factor of 2.  
We also use the data listed in Table 4, which have been  smoothed to the size of the background continuum source rather
than to the spatial resolution of the SWAS beam.
 The densities and column densities determined from this data set, however, do 
not differ significantly from an identical analysis using data coadded over an entire SWAS beam.

We created a $20\times 20$ grid of models in density-column density parameter space 
using a constant kinetic temperature of 8 K (Vastel et al. 2000), consistent with what is derived from
our $^{12}$CO observations assuming that the emission is optically thick and thermalized.
 The densities range from 
$1000 - 5000 $ cm$^{-3}$ and 
the $^{13}$CO column densities (per velocity interval) range from $10^{13} - 10^{16}$ cm$^{-2}$/km s$^{-1}$.  
To fit the observations of the $^{13}$CO and C$^{18}$O isotopologues we assume isotopic 
abundance ratios (with respect to CO) of 
55 and 500 respectively.  The observed line intensities are fit to the grid of LVG models using a 
$\chi^2$ minimization routine to find the density and column density combination that best fit the 
observations.  
Results from the LVG fitting are listed  in Table 6 along with the estimated visual extinctions where 
we have assumed that  N(H$_2$) $ \sim 8 \times 10^{20} A_V$ and N(H$_2$)/N(CO) $\sim 10^4$.  It is important to note that this only estimates the dust extinction in the CO emitting 
region and ignores any additional extinction which may arise in overlying layers of HI or H$_2$.
To check the sensitivity of our results to our assumed value of kinetic temperature, we performed an identical LVG analysis using kinetic temperature of
15 K instead of 8 K.  The resultant  densities change by less than a factor
of three and the column densities by less than 25\%.

Using the H$_2$ density in Table 6  we also calculate the neutral carbon (C$^o$) column 
density from the SWAS [C~I] $^3P_1 \rightarrow ^3P_0$ observations (using the same LVG code 
and again assuming a kinetic temperature of 8 K).  The results are also tabulated in Table 6 along 
with the C$^o$/CO abundance ratio.  The C$^o$/CO abundance ratio varies from 0.6 for clouds with inferred
visual extinctions (A$_V$) of 
1.9, to C$^o$/CO = 3.8 for clouds with inferred A$_V$ = 0.1.  These numbers are consistent with C$^o$/CO  ratios seen at the low 
column density edges of UV illuminated giant molecular clouds (e.g. Plume et al 1999) and with 
previous observations of  high-latitude and translucent 
clouds (Ingalls et al 1997; Stark et al 1996; Stark \& van Dishoeck 1994).  

\subsection{H$_2$O and OH Abundances}

The SWAS ortho-H$_2$O  column densities are derived from a ``curve of growth''
analysis of SWAS o-H$_2^{18}$O and o-H$_2$O spectra.  The column density in the lower state 
($N_l$)can be expressed as:
$$
N_l = {{g_l}\over{g_u}}{{8\pi \nu_o^3}\over{c^3}}{{\sqrt{\pi}}\over{A_{ul}}}{{1}\over{[1-e^{-h\nu/kT_{ex}}]}}\tau _o {{\Delta V(FWHM)}\over{2\sqrt{ln 2}}},~~~~~~(1)
$$
where $\tau_o$ is the optical depth at line center.  For the o-H$_2$O and o-H$_2^{18}$O absorption lines in W49, T$_{ex}$ is probably much less than 
$h\nu /k$ = 27 K, and so equation (1) reduces to:
$$
N_l = {{g_l}\over{g_u}}{{8\pi}\over{\lambda _o^3}}{{\sqrt{\pi}}\over{A_{ul}}}~~\tau _o {{\Delta V(FWHM)}\over{2\sqrt{ln 2}}}.~~~~~~(2)
$$

Using parameters 
appropriate for the SWAS  observations: $\lambda (o-H_2^{18}O) = 547.39 \mu m$, $\lambda (o-H_2O) = 538.66 \mu m$,
$A_{ul} = 3.5 \times 10^{-3}$ s$^{-1}$ (for both o-H$_2$O and o-H$_2^{18}$O), and $g_l/g_u = 1$, the expressions for
the column density reduce to:
$$
N_l(o-H_2^{18}O) = 4.66 \times 10^{12} ~~\tau _o(H_2^{18}O) \Delta V(FWHM),~~~~~(3)
$$
and
$$
N_l(o-H_2O) = 4.89 \times 10^{12}  ~~\tau _o(H_2O) \Delta V(FWHM),~~~~~(4)
$$
where $\Delta V(FWHM)$ is in units of km s$^{-1}$.  Again, since the excitation temperatures in 
these clouds is low, $N_l$ is, to first approximation, equal to the total column density.  $\tau_o$ is 
the line center optical depth of the particular molecular transition.

Table 7 presents the ortho-water column density in each of the six line-of-sight
absorption features towards W49A.  The first row lists upper limits on $\tau _o \Delta V$ for o-H$_2^{18}$O taken from the
o-H$_2^{18}$O observations listed in Table 2.  The second row presents upper limits on the o-H$_2^{18}$O column 
densities as calculated from
equation (3). The third row gives the o-H$_2$O column density upper limit calculated
under the assumption that $N(H_2O) = 500 \times N(H_2^{18}O)$.  The fourth row lists  $\tau _o \Delta V$ for 
o-H$_2$O taken from the o-H$_2$O observations listed in Table 2.  The fifth row presents the o-H$_2$O column densities 
as calculated from equation (4).  The sixth row lists the CO column density derived from  our 
observations of CO and its isotopologues.
The seventh row of Table 7 presents the o-H$_2$O abundance as determined from the 
o-H$_2^{18}$O observations (i.e. by dividing row 3 by row 6) and assuming that the CO abundance 
(relative to H$_2$) is 10$^{-4}$.  Row eight provides the same o-H$_2$O abundance except that in this 
case we use the o-H$_2$O observations directly.  To obtain the total H$_2$O 
abundances, the numbers listed in Table 7 must be multiplied by 4/3 to account for the assumed ortho to para ratio.

The  o-H$_2^{18}$O observations allow us to obtain upper limits for the o-H$_2$O abundance, whereas the o-H$_2$O  observations set lower limits.  Thus, 
by using both the o-H$_2^{18}$O and o-H$_2$O observations, we are able to constrain the o-H$_2$O 
abundances  in each line-of-sight cloud to within about an order of magnitude.  Table 7 shows that the water abundance in the three clouds with optically thin water lines is between $8.1 \times 10^{-8}$ and $4 \times
10^{-7}$ .  In the three additional clouds, where the H$_2$O lines are saturated, the upper limits to the water abundance are between $8.2\times 10^{-8}$ and $1.5\times10^{-6}$.
 These abundances are similar to those determined for a line-of-sight ortho-water absorption 
feature towards W51 (Neufeld et al. 2002),  Sgr B2 (Neufeld et al. 2000; Cernicharo et al. 1997), and Sgr A* (Moneti et al. 2001).  
The water abundances in these absorption features are, however, higher than those seen from ortho-water
 {\it emission} lines in other giant molecular clouds ($2 - 10 \times 10^{-9}$; Snell et al. 2000b).  An explanation for the low water
 abundances observed in typical giant molecular clouds observed by Snell et al. (2000b) is given by Bergin et al. (2000) who suggest that, in the dense interiors of these well-shielded cores, H$_2$O can readily freeze out onto dust grains.

To calculate the OH abundances, we again use equation(1).  However, for OH, $h\nu < kT_{ex}$ 
even in the cold absorbing clouds and, therefore:
$$
N_l = {{g_l}\over{g_u}}{{8\pi \nu_o^2 k}\over{h c^3}}{{\sqrt{\pi}}\over{A_{ul}}} ~T_{ex} ~\tau _o {{\Delta V(FWHM)}\over{2\sqrt{ln 2}}}.~~~~~~(5)
$$
Inserting the appropriate constants for the 1665 MHz ($A_{1665} = 7.11\times 10^{-11}$ 
and $g_l/g_u = 3/3 = 1$) and the 1667 MHz lines
($A_{1667} = 7.71\times 10^{-11}$ and $g_l/g_u = 5/5 = 1$) we obtain the following equations for column density:
$$
N_{tot}(1665)= {{16}\over{3}} \times N_l(1665) = 4.30 \times 10^{14} ~ T_{ex} ~ \tau _o(1665) \Delta V(FWHM),~~~~~~(6)
$$
and
$$
N_{tot}(1667)= {{16}\over{5}} \times N_l(1667) = 2.38 \times 10^{14} ~ T_{ex} ~ \tau _o(1667) \Delta V(FWHM),~~~~~~(7)
$$
where again $\Delta V(FWHM)$ is in units of km s$^{-1}$.  The factors 16/3 and 16/5 account for the relative populations in the four OH  hyperfine levels.

To determine the OH column densities, therefore, we need to know the excitation temperature.  
Since the lines at 54.2 and 68.8 \kms\ appear to be in LTE, we assign $T_{ex}$ = 8 K as we
did for the CO analysis.  The other four clouds, however, appear to suffer significantly from non-LTE excitation and, since we have no way of knowing the excitation temperature in these clouds, we cannot accurately calculate their column densities and do not consider them in the following discussion.   Fortunately, the 54.2 and 68.8 \kms\ clouds are also those in which the H$_2$O lines are optically thin, and so provide us with our best measure of the water column density.  
Table 8 shows that the OH column densities in the remaining two clouds (where N(OH) is the average of the 1665 and 1667 MHz results) range from $3.5 \times 10^{13}$ cm$^{-2}$ to $1.4 \times 10^{14}$ cm$^{-2}$, similar to the value of 
$8\times10^{13}$ cm$^{-2}$ seen in the line of sight cloud towards W51 (Neufeld et al. 2002).

\section{Discussion}

\subsection{H$_2$O and OH Abundances: A New Perspective on Branching Ratios}

The H$_2$O/OH abundance ratio can provide insight into the chemical networks that produce oxygen-bearing molecules.  Recently, Neufeld et al. (2002) made predictions for the ratio of OH and H$_2$O column 
densities by using a simple analytic model  in which OH and H$_2$O formation, 
through dissociative recombination of H$_3$O$^+$ (producing either O, OH, or H$_2$O), was balanced by photodissociation.   Neufeld (2002) compared the results of the simple analytic model with 
the detailed results of a diffuse cloud PDR model (e.g. Kaufman et al. 1999) and found that the simple analytic 
model predictions were robust for clouds of gas density $n=100$ cm$^{-3}$, for a range of cloud
extinctions and cosmic ray ionization rates.  Those model results
show that cosmic ray ionization of H$_2$ leads directly to the formation of H$_3$O$^+$ which is subsequently converted 
to H$_2$O and OH through dissociative recombination. 

The OH and H$_2$O abundances, however, depend sensitively upon the poorly constrained chemical branching
ratios ($f_O$, $f_{OH}$, and $f_{H_2O}$) for dissociative recombination of H$_3$O$^+$. 
The flowing afterglow  laboratory experiment by  Williams et al. (1996) found  $f_{\rm H_2O}$:$f_{\rm OH}$:$f_{\rm O}$ = 0.05:0.65:0.3.  Two other experiments using a different  technique, one using the ASTRID heavy-ion storage ring (Vejby-Christensen et al. 1997; Jensen et al. 2000) and the other using the CRYRING heavy-ion storage ring (Neau et al. 2000), yielded  $f_{\rm H_2O}$:$f_{\rm OH}$:$f_{\rm O}$ = $0.25\pm 0.01:0.74\pm 0.02:0.013\pm 0.005$ and $f_{\rm H_2O}$:$f_{\rm OH}$:$f_{\rm O}$ = $0.18\pm 0.05:0.78\pm 0.08:0.03\pm 0.06$ respectively.
Observations of diffuse clouds in the ISM have yielded similarly discrepant results.  
Recent UV observations with HST towards the moderately 
reddened star HD 154368  (Spaans et al. 1998) found a
$3\sigma$ upper limit on $f_{H_2O}$ of 0.06, which is consistent with the flowing afterglow experiment. 
However, observations of a line of sight cloud seen in absorption against W51's continuum 
are, instead, consistent with the ASTRID and CRYRING experiments ($f_{H_2O} \sim 0.25$; Neufeld et al. 2002).  Our results, however, are consistent with {\it neither} of the experimental values for the branching ratios.
In Figure \ref{fig:branch}
we show the results of the simple analytic model for branching ratios $f_{\rm H_2O}$:$f_{\rm OH}$:$f_{\rm O}$  
of 0.05:0.65:0.30  and 0.25:0.75:0.0 (the top and bottom dashed lines respectively). We also show the measured values (or upper/lower limits)
for two of the six observed absorption features (for which we have OH column densities). As may be seen in Figure \ref{fig:branch}, the observations
lie below and to the right of the line relevant for the branching ratio 0.25:0.75:0.0, and are even less consistent with the smaller H$_2$O branching ratio, at least
in light of the conclusions of the simple analytic model.

Neufeld et al. (2002) provide an explanation for the discrepancy in the branching ratios.   When formation
of H$_2$O and OH via dissociative recombination of H$_3$O$^+ $ is the {\it only} formation route, 
the predicted ratio of column densities depends only on the branching ratios. 
If, however, there are significant other routes to 
OH or H$_2$O, then the ratio of column densities will depart from the simple analytic model predictions. 
If even a small fraction of gas in the clouds is warm (above $\sim$ 300 K) then neutral-neutral reactions, which
have relatively high activation barriers,  can begin to contribute to the production of OH and H$_2$O.  Neufeld et al. (2002) show that,  by varying the temperature of the gas, it is possible to match either of the experimental branching ratios.  This model, however, requires a small fraction of gas with temperatures in excess of 600 K to explain the OH and H$_2$O column densities in the clouds observed towards W49 (Figure \ref{fig:branch}).  In the remainder of this section, we present an additional explanation for the observed OH and H$_2$O column
densities which builds on the Neufeld et al. (2002) results.

To investigate additional chemical effects on the column density ratio, we have run models similar to those
in the Neufeld et al. (2002) paper, but extending to higher gas densities. In all of the models, we assume that a 
diffuse cloud is illuminated from both sides by a total interstellar field equal to G$_0$=1.7 which corresponds to 
the current best estimate of the local
interstellar radiation field (a value G$_0$=1 corresponds to the Habing (1968) UV field; $1.3\times 10^{-4}$ 
erg cm$^{-2}$ s$^{-1}$ sr$^{-1}$). 
In accordance with the densities and column densities derived from our LVG analysis (Table 6), 
we compute the total water and hydroxyl column densities through clouds with gas densities
$n=100, 10^3$ and $10^4$ cm$^{-3}$ and with total visual extinctions of A$_V$ = 1, 2, 3, and 4. 
The gas temperature is solved for self-consistently and, at the center of the clouds, is found to be  22 K for A$_V$ = 2, 12 K for A$_V$ = 3, and 8 K
for A$_V$ = 4.
These values of $A_V$ are consistent with those estimated from our $^{13}$CO observations 
since they are the {\it total} $A_V$ through the cloud, whereas the values listed in Table 6 only 
consider the  CO emitting region. 
In the surface
layers (to $A_V \sim 1$ on each side of the cloud) there is essentially no CO (the CO/H$_2$ abundance ratio $< 10^{-5}$) but the
dust in these H and H$_2$ layers still contributes
to the overall extinction.  Therefore, one needs to add approximately 2 magnitudes of extinction to the values listed in Table 6 to compare with our models.

The destruction of OH and H$_2$O need not be dominated by
    photodestruction as in the simple model of Neufeld et al (2002). 
    OH can be destroyed by neutral-neutral reactions with atomic O.
    The photodestruction rate per OH molecule is proportional to G$_0$,
    which is held fixed, but the destruction by O is proportional
    to $n$(O), the density of atomic O.  Thus, at higher $n$, the latter
    mechanism dominates, OH is destroyed more rapidly and the
    H$_2$O/OH column density ratio increases relative to the simple toy 
    model.  Similarly, at high extinction, the photodestruction rate
    decreases due to dust attenuation of the FUV field, but the neutral
    rate is unaffected, leading to a relatively higher rate of destruction
    of OH than H$_2$O and an increased H$_2$O/OH ratio.
    These effects can be seen in Figure  \ref{fig:branch} which shows an increase in the ratio of H$_2$O  to OH column density  as the gas density and visual extinction increase.
 The fact that all four features seen in W49 lie at
    or below the $f_{H_2O}$ = 0.25 curve (and much below the  $f_{H_2O}$ =
    0.05 curve) suggests both that we can rule out a branching ratio of 
     $f_{H_2O} =0.05$,
    and that a simple model which incorporates only dissociative recombination 
is inadequate.

Based on ISO observations of the [O~I] 63 \um\ transition, Vastel et al. (2000) have suggested 
that CO is depleted by at least a factor of 6 in these clouds.  If true, then our H$_2$O and OH abundances 
would also need to be lowered by a similar amount.  However, there are several potential difficulties with this interpretation.  First, while the authors analysed their data as carefully as possible, it is inherently difficult to compare the 63 $\mu$m [O~I] absorption feature to the HI and molecular features due to the poor spectral resolution  of ISO ($\Delta V_{FWHM} \sim 44$ \kms ).  Second, the HI column densities quoted by Vastel et al (2000) are a $few \times 10^{21}$ 
cm$^{-2}$ and the HI is fairly optically thick ($\tau \sim 2-4$; Lockhart \& Goss 1978).  Therefore, it is possible
that they are underestimating the HI column density, in which case the atomic gas may account for more of the
observed [O~I] 63 $\mu$m absorption.  Finally,  if CO is freezing out on grains, then oxygen must suffer the same fate.  Calculations by Bergin et al. (2000) show that atomic oxygen depletes onto grains even more readily than CO.   
In fact, in the line-of-sight clouds towards W49, our pure gas-phase models are able to produce results that are consistent with the observed H$_2$O/CO column density ratios, without having to resort to freeze-out of CO molecules onto dust grains.
In figure \ref{fig:h2oco} we present the results of our PDR calculations along with the observed H$_2$O and CO column densities. 
The  H$_2$O/CO  abundance ratio in the three clouds with optically thin water lines is between $8.1 \times 10^{-4}$ and $4 \times
10^{-3}$ and, in the three clouds where the H$_2$O lines are saturated, the upper limits to the H$_2$O/CO  ratio is between $8.2\times 10^{-4}$ and $1.5\times10^{-2}$ (Table 7). 
 Figure  \ref{fig:h2oco} shows that the models are consistent with the observations for A$_V \ge 3$ and n $\ge 100$ cm$^{-3}$.  

If depletion does play a role in these clouds, then the freeze-out of atomic oxygen may help to explain the observed H$_2$O/OH abundance ratios in clouds with pressures more closely matched to those observed in other diffuse interstellar clouds.  
For example, the 54  km s$^{-1}$ feature is consistent with models having A$_V$ between 3 and 4 and intermediate densities (n $< 10^4$ cm$^{-3}$).   The  68 km s$^{-1}$ feature, however, requires higher density models. The resultant pressures
($nT\sim few\times 10^5\,\,cm^{-3}\,K$) in this cloud, therefore, is quite high for a diffuse interstellar cloud. 
However, it is possible to
lower the gas density and still obtain high H$_2$O/OH column density ratios if we 
include chemical reactions on 
the surfaces of grains. Preliminary calculations indicate that the inclusion of grain chemistry, involving O and C bearing
species, leads naturally to the observed high H$_2$O-to-OH column density ratios.
 In these
    models, photodesorption of water ice formed on grain surfaces is
    the main source of gas phase H$_2$O  throughout the clouds (atomic oxygen sticks to the grains and rapidly is converted to water
    ice which is photodesorbed).  This results in a higher water column density than found in the
models without grain chemistry. 
OH, on the other hand continues to be formed primarily via the dissocative recombination
of H$_3$O$^+$. The point labeled as ``grain" in Figure {\ref{fig:branch} shows the effect of adding grain surface chemistry even in very low density gas ($n\sim 10^2\,\rm cm^{-3}$).  Although the inclusion of grain surface chemistry is not needed to explain the H$_2$O/OH ratios in the clouds towards W49, the attractive feature of this model is that we can still produce high H$_2$O-to-OH column density ratios in lower 
density gas (n $< 10^4$ cm$^{-3}$) with pressures more reflective of diffuse clouds. 
We will explore the effects of grain chemistry in diffuse clouds in more detail in a subsequent paper. 

\subsection{CO \& [CI] Intensities}

We have also used our models to predict the strengths of individual [C~I] (492 GHz) and $^{13}$CO ~$J=1 \rightarrow 0$ emission lines.
Since  [C~I] is observed
in emission with a relatively large beam, we need some way to estimate the emission produced by the same gas that produced
the H$_2$O and OH absorption features. To do this we compare the $^{13}$CO intensity averaged over the $\sim
1.5\arcmin$ continuum source (Table 4) to
 that averaged over the SWAS beam size; the ratio of these two intensities is then used to scale the observed
[C~I] emission in order to estimate how much of it comes from the direction of the continuum source. Corrected values for [C~I] are given in Table 3.

In Figure  \ref{fig:cico}, we plot the observed (corrected) [C~I] and $^{13}$CO integrated intensities and the intensities predicted from
the standard PDR models. 
The uncorrected  [C~I] and $^{13}$CO integrated intensities smoothed to the resolution of the SWAS [C~I] observations are not shown, but do not differ significantly from the points plotted in  Figure  \ref{fig:cico}.
The observed intensities are not well matched by the models. For instance, the models can 
only match the observed CO intensities by resorting to high gas densities, a known problem with PDR models noted by other 
authors (e.g. Bensch et al. 2003). These high density models, however, overpredict the [C~I] intensity. There are a variety of ways 
in which this shortcoming could be resolved. For instance, if CO self-shielded more effectively than is generally assumed,
conversion of C$^o$ to CO would happen closer to the cloud surface. Then the CO intensity would be higher and the [C~I] intensity 
would be lower. Another solution might be the use of constant pressure PDR models; our models 
assume that the gas
density is constant. If the [C~I] emission came from gas with a density below the critical density for the 492 GHz transition, while
CO emission came from higher density (though cooler) gas, then the desired effect might also be achieved. However, in a preliminary 
constant pressure calculation with low surface density and low UV field, 
the density contrast between the [C~I] and CO regions
was only a factor $\sim 2$; not enough to explain the observed differences. The discrepancies between the [C~I] and \tco\ observed and model line strengths will be investigated more thoroughly in a future paper.

\section{Conclusions}

We have analysed emission and absorption lines from 6 clouds along the line-of-sight toward W49A.  
Using the Submillimeter Wave Astronomy Satellite (SWAS), we observed
[C~I] $^3P_1 \rightarrow ^3P_0$ ($\nu = 492.1607$ GHz), 
$^{13}$CO ~$J = 5\rightarrow 4$    ($\nu = 550.926$ GHz), 
o-H$_2$O $1_{10}\rightarrow 1_{01}$ ($\nu = 556.936$ GHz) and o-H$_2^{18}$O $1_{10}\rightarrow 1_{01}$ transition ($\nu = 547.676$ GHz).  
We observed $^{12}$CO ~$J = 1\rightarrow 0$ ($\nu = 115.271$ GHz),  $^{13}$CO ~$J = 1\rightarrow 0$ 
($\nu = 110.201$ GHz), and $C^{18}O ~J = 1\rightarrow 0$ ($\nu = 109.782$ GHz),
at the 14m Five College Radio Astronomy Observatory (FCRAO) and the $^{12}$CO ~$J = 3 \rightarrow 2$ transition  at
the KOSMA (K\"olner Observatorium f\"ur Submm-Astronomie).  We also observed the 1665 and 1667 
MHz transitions of OH at the Arecibo Observatory, and mapped the 450$\mu$m continuum emission at the Caltech Submillimeter
Observatory (CSO).

 The o-H$_2$O  and OH (1665 and 1667 MHz) transitions are observed in {\it absorption}, whereas the 
 other lines are seen in {\it emission} (with the exception of o-H$_2^{18}$O and $^{13}$CO ~$J = 5 \rightarrow 4$ 
 which were not detected).  Using the emission lines of $^{13}$CO and C$^{18}$O 
 we derive gas densities of $1500 - 3000$ cm$^{-3}$ and CO column densities of 
 $\sim 7.9 \times 10^{15} - 2.8 \times 10^{17}$ cm$^{-2}$ via a standard Large Velocity Gradient analysis.   
 The observations of o-H$_2$O and OH {\it absorption} have the advantage that their
column densities can be
derived without  knowledge of
the temperature or density of the absorbing gas.  
By using both the o-H$_2^{18}$O and o-H$_2$O absorption lines, we are able to constrain the column-averaged o-H$_2$O 
abundances  in each line-of-sight cloud to within about an order of magnitude.  
We find  N(o-H$_2$O)/N(H$_2$) = $8.1 \times 10^{-8} - 4 \times
10^{-7}$ for three clouds with optically thin water lines.  In three additional clouds where the H$_2$O lines are saturated, we have used observations of the H$_2^{18}$O ground-state transition to find  upper limits to the water abundance of $8.2\times 10^{-8} - 1.5\times10^{-6}$.
The o-H$_2$O abundances are similar to those determined for a line-of-sight water absorption 
feature towards W51 and Sgr B2 (Neufeld et al. 2000; 2002) but are higher than those seen from water
 {\it emission} lines in molecular clouds (Snell et al. 2000b).    
 We measure the OH abundance  from the average of the 1665 and 1667 MHz observations and find   N(OH)/N(H$_2$) = $1.2 \times 10^{-7} - 1.1 \times 10^{-6}$.

If dissociative recombination is the primary formation pathway for OH and H$_2$O then the abundances of these 2 species depends sensitively on the branching ratios ( $f_{\rm H_2O}:f_{\rm OH}:f_{\rm O}$).
Observations and theoretical work has set  these branching ratios to fairly discrepant values of either
0.25:0.75:0 or  0.05:0.65:0.3.  However, based on a simple analytical model, none of the features seen in W49 appear to be consistent
with either value of the branching ratio.
This suggests that a simple model which incorporates only dissociative recombination is inadequate, and that additional chemical effects need to be considered.
Building on the work of Neufeld et al. (2002),
our photo-chemistry models 
provide an additional explanation for the observed OH and H$_2$O column
densities   by including depth-dependent photodissociation, neutral-neutral reactions which preferentially destroy the OH.  The photo-chemistry models
can explain the observed OH and H$_2$O column densities if  $f_{H_2O} =0.25$ but not if
     $f_{H_2O} =0.05$.  These gas-phase models are also able to produce results that are consistent with the observed H$_2$O/CO column density ratios, without having to resort to freeze-out of CO molecules onto dust grains. However, it is possible that  atomic oxygen can stick to the grains and rapidly converted to water
    ice which is photodesorbed.  One attractive feature of this model is that we can still produce high H$_2$O-to-OH column density ratios in lower 
density gas (n $< 10^4$ cm$^{-3}$)
 with pressures that more closely match those observed in other diffuse clouds.

\acknowledgements
This work was supported by NASA Grant NAS5-30702 (to SWAS).  R.P was also supported through a 
grant from the Natural Sciences and Engineering Research Council of Canada.
The CSO is funded by NSF contract AST 96-15025.  The National Astronomy and Ionosphere Center 
is operated by Cornell University under a Cooperative
Agreement with the National Science Foundation.  The Five College Radio
Astronomy Observatory is operated with the permission of
the Metropolitan District Commission, Commonwealth of
Massachusetts, and with the support of the National Science
Foundation under grant AST 01-00793.

\center{
\begin{table*}
\caption{Spectral Line Observations of W49A} 
\vskip 0.3 true in
 
\begin{tabular}{lccc}
\tableline\tableline
Telescope & Line & On Source Int. Time & $1\sigma$ rms  \\
       &      &   (hours)           & (mK) \\
\tableline
 
SWAS  & [C~I] $^3P_1 \rightarrow ^3P_0$        & 405   & 5    \\
...   & $^{13}$CO ~$J = 5\rightarrow 4$           & 63    & 8  \\
...   & H$_2$O $1_{10}\rightarrow 1_{01}$      & 63    & 8   \\
...   & H$_2^{18}$O $1_{10}\rightarrow 1_{01}$ & 342   & 5    \\
FCRAO & $^{12}$CO ~$J = 1\rightarrow 0$           &  3    & 50  \\
...   & $^{13}$CO ~$J = 1\rightarrow 0$           &  3.2  & 14  \\
...   & C$^{18}$O ~$J = 1\rightarrow 0$           &  4    & 13  \\
KOSMA & $^{12}$CO ~$J = 3\rightarrow 2$           &  0.5    & 60  \\
Arecibo  & OH 1665 MHz                   &  0.17 & 517 \\
...             & OH 1667 MHz                   &  0.17 & 673 \\

\tableline\tableline \\
 
\end{tabular}
\end{table*}
}

\center{
\begin{table*}
\caption{Gaussian Fit Parameters for the Line-of-Sight Absorption Features:H$_2$O \& H$_2^{18}$O} 
\vskip 0.3 true in
 
\begin{tabular}{lcccccc}
\tableline\tableline
Line & $\Delta T_A^*$~$^1$  & $V_{LSR}$     & $\Delta V_{FWHM}$ & $\int{\Delta T_A^*dV}$ & $F_l/F_c$ & $\tau_o$ \\
     & (K)      & (km s$^{-1}$) & (km s$^{-1}$)     & (K km s$^{-1}$) &  & \\
\tableline
 
H$_2$O $1_{10}\rightarrow 1_{01}$      & -0.067   & 33.5 & 3.5 & -0.25  & 0.22 & 1.9 \\
...                                    & -0.080   & 39.5 & 4.1 & -0.35 & 0.06 & $>$ 2.7 \\
...                                    & -0.051   & 53.5 & 6.0 & -0.33 & 0.40 & 1.0 \\
...                                    & -0.085   & 59.6 & 4.0 & -0.37 & 0.006 & $>$ 5.1 \\
...                                    & -0.073   & 63.3 & 2.7 & -0.21 & 0.15& $>$1.9 \\
...                                    & -0.032   & 68.0 & 5.0 & -0.17 & 0.63 & 0.5 \\

H$_2^{18}$O $1_{10}\rightarrow 1_{01}$ & $<$ -0.012 & 34.0 & 4.6 & $<$ -0.059  & $>$ 0.86 & $<$ 0.15 \\
...                                    & $<$ -0.006 & 39.8 & 3.2 & $<$ -0.020  & $>$ 0.93 & $<$ 0.07\\
...                                    & $<$ -0.003 & 54.0 & 5.0 & $<$ -0.015 & $>$ 0.97 & $<$ 0.04  \\
...                                    & $<$ -0.007 & 59.5 & 3.4 & $<$ -0.025 & $>$ 0.92 & $<$ 0.09 \\
...                                    & $<$ -0.003 & 63.3 & 2.7 & $<$ -0.010 & $>$ 0.97 & $<$ 0.04 \\
...                                    & $<$ -0.003 & 69.0 & 5.1 & $<$ -0.015 & $>$ 0.97 & $<$ 0.04 \\
 
\tableline\tableline \\
 
\end{tabular} \\
$1 - \Delta T_A^*$ denotes the baseline subtracted antenna temperature which, for absorption lines, is a negative quantity.
\end{table*}
}

\center{
\begin{table*}
\caption{Gaussian Fit Parameters for the Line-of-Sight Emission Features: [C~I] } 
\vskip 0.3 true in
%
%

%

\begin{tabular}{lccccc}
\tableline\tableline
Line & $\Delta T_A^*$  & $V_{LSR}$     & $\Delta V_{FWHM}$ & $\int{\Delta T_A^*dV}$  & $\int{\Delta T_A^*dV}$(corrected)$^1$\\
     & (K)      & (km s$^{-1}$) & (km s$^{-1}$)     & (K km s$^{-1}$) & (K km s$^{-1}$) \\
\tableline
 
[C~I] $^3P_1 \rightarrow ^3P_0$        & 0.11     & 34.0 & 4.6 & 0.56 & 1.00 \\
...                                    & 0.45     & 39.8 & 3.2 & 1.52 & 1.26 \\
...                                    & 0.13     & 54.0 & 5.0 & 0.71 & 0.65 \\
...                                    & 0.36     & 59.5 & 3.4 & 1.35 & 0.66 \\
...                                    & 0.62     & 63.3 & 2.7 & 1.76 & 2.90 \\
...                                    & 0.12     & 69.0 & 5.1 & 0.70 & 0.44  \\

\tableline\tableline \\
 
\end{tabular}\\
$^1$ - Corrected to a a 1.5$'$ beam (see Section 4.2).
\end{table*}
}

\center{
\begin{table*}
\caption{Gaussian Fit Parameters for the Emission Features: CO Isotopologues} 
\vskip 0.3 true in
 
\begin{tabular}{lcccc}
\tableline\tableline
Line & $\Delta T_A^*$  & $V_{LSR}$     & $\Delta V_{FWHM}$ & $\int{\Delta T_A^*dV}$ \\
     & (K)      & (km s$^{-1}$) & (km s$^{-1}$)     & (K km s$^{-1}$) \\
\tableline

$^{12}$CO ~$J = 1\rightarrow 0$~$^a$           &  0.41    & 33.5 & 2.0 & 0.88  \\
...                                    &  1.43    & 39.3 & 3.0 & 4.57  \\
...                                    &  0.33    & 54.0 & 3.7 & 1.33  \\
...                                    &  0.76    & 59.0 & 3.3 & 2.62  \\
...                                    &  1.47    & 63.0 & 2.9 & 4.64  \\
...                                    &  0.36    & 69.0 & 2.4 & 0.94  \\

$^{12}$CO ~$J = 3\rightarrow 2$           &  1.42    & 39.3 & 3.0 & 4.61  \\
...                                    &  0.24    & 54.0 & 3.5 & 0.88  \\
...                                    &  0.59    & 58.5 & 3.3 & 2.08  \\
...                                    &  1.49    & 63.0 & 2.7 & 4.22  \\
...                                    &  0.31    & 69.0 & 5.1 & 1.68  \\

$^{13}$CO ~$J = 1\rightarrow 0$~$^a$           &  0.03    & 33.5 & 2.5 & 0.09   \\
...                                    &  0.68    & 39.4 & 1.7 & 1.23  \\
...                                    &  0.02    & 53.5 & 6.0 & 0.11  \\
...                                    &  0.13    & 59.4 & 3.2 & 0.45  \\
...                                    &  0.96   & 63.1 & 1.5 & 1.52 \\
...                                    &  0.03    & 68.7 & 4.6 & 0.15  \\

$^{13}$CO ~$J = 5\rightarrow 4$           & $<$0.008 &  -   &  -  &  -   \\

C$^{18}$O ~$J = 1\rightarrow 0$~$^a$           &  0.09    & 39.4 & 1.0 & 0.09  \\
...                                    &  0.18    & 63.1 & 1.0 & 0.19  \\

\tableline\tableline \\
 
\end{tabular} \\
$^a$ - data smoothed to 1.5$'$ resolution.
\end{table*}
}

\center{
\begin{table*}
\caption{Gaussian Fit Parameters for the Line-of-Sight Absorption Features: OH} 
\vskip 0.3 true in
 
\begin{tabular}{lcccccc}
\tableline\tableline
Line & $\Delta T_A^*$  & $V_{LSR}$     & $\Delta V_{FWHM}$ & $\int{\Delta T_A^*dV}$ & $F_l/F_c$ & $\tau_o$\\
     & (K)      & (km s$^{-1}$) & (km s$^{-1}$)     & (K km s$^{-1}$) &  & \\
\tableline
 
OH 1665 MHz              & -1.3     & 33.6 & 3.5 & -4.8    & 0.995 & 0.005 \\
...                                    & -8.3      & 39.3 & 1.7 & -15.4 & 0.967 & 0.034 \\
...                                    & -1.9      & 54.2 & 4.7 & -9.7    & 0.992 & 0.008 \\
...                                    & -3.9      & 59.8 & 3.5 & -14.5 & 0.984 & 0.016 \\
...                                    & -14.2    & 62.9 & 1.8 & -27.1 & 0.943 & 0.059 \\
...                                    & -1.0      & 68.8 & 2.8 & -3.0    & 0.996 & 0.004  \\

OH 1667 MHz              & -3.9     & 33.6 & 3.4 & -14.3 & 0.985 & 0.016 \\
...                                    & -8.1      & 39.3 & 2.4 & -20.8 & 0.968 & 0.033 \\
...                                    &  -3.9     & 54.2 & 4.7 & -19.6 & 0.985 & 0.016 \\
...                                    & -12.0    & 59.8 & 2.9 & -37.6 & 0.952 & 0.049 \\
...                                    & -24.6    & 62.9 & 1.8 & -46.4 & 0.902 & 0.103 \\
...                                    & -1.5      & 68.8 & 2.8 & -4.5   & 0.994 & 0.006  \\

\tableline\tableline \\
 
\end{tabular}
\end{table*}
}

\center{
\begin{table*}
\caption{LVG Model Results} 
\vskip 0.3 true in
 
\begin{tabular}{lcccccc}
\tableline\tableline
V$_{LSR}$     & log(n$_{H_2}$)           & log(N($^{13}$CO))    & A$_V^b$ & log(N(C$^o$))  & N(C$^o$)/N($^{12}$CO)$^a$  \\
 (km s$^{-1}$)  &  (cm$^{-3}$)  &  (cm$^{-2}$) & (mag) & (cm$^{-2})$ &  \\
\tableline
 
33.5     & 3.48 & 14.16  & 0.1 & 16.48 & 3.8 \\
39.4     & 3.52 & 15.44  & 1.9 & 16.99 & 0.6 \\
53.5     & 3.48 & 14.37  & 0.2 & 16.59 & 3.0 \\
59.4     & 3.48 & 14.92  & 0.6 & 16.91 & 1.8 \\
63.1     & 3.18 & 15.70  & 3.4 & 17.20 & 0.6\\
68.7     & 3.48 & 14.45  & 0.2 & 16.56 & 2.3 \\

\tableline\tableline \\
 
\end{tabular}
\\
$a$ Assuming N($^{12}$CO):N($^{13}$CO) = 55:1. \\
$b$ The visual extinction in the CO emitting layer.  Add approximately 2 magnitudes to get the {\it total} A$_V$.
\end{table*}
}

 \center{
\begin{table*}
\caption{Estimated Water Column Densities and Abundances in W49A} 
\vskip 0.3 true in
\hskip -0.3 in
\scriptsize{ 
\begin{tabular}{lcccccc}
\tableline\tableline
   &      &      & $V_{LSR}$ (\kms) &      &      &       \\
   & 33.5 & 39.5 &     53.5         & 59.6 & 63.3 & 68.0  \\
\tableline

$\tau_o(o-H_2^{18}O)\Delta V$ (km s$^{-1}$)        & $< 0.69 $ & $< 0.23$  & $< 0.18$ & $< 0.29$ & $< 0.10$ & $< 0.18$ \\

$N(o-H_2^{18}O)$  (cm$^{-2}$)            & $<  3.2 \times 10^{12}$ & $< 1.1 \times 10^{12}$ & $< 8.4 \times 10^{11}$ & $< 1.3 \times 10^{12}$ & $< 4.5 \times 10^{11}$ & $< 8.6 \times 10^{11}$ \\

$N(o-H_2O)~^a$    (cm$^{-2}$)               & $< 1.6 \times 10^{15}$ & $< 5.4 \times 10^{14}$ & $< 4.2 \times 10^{14}$ & $< 6.7 \times 10^{14}$ & $< 2.3 \times 10^{14}$ & $< 4.3 \times 10^{14}$ \\

$\tau_o(o-H_2O)\Delta V$ (km s$^{-1}$)    & 6.5 & $>$11.3  & 6.2 & $>$ 20.6 & $>$ 5.2 & 2.6  \\

$N(o-H_2O)$    (cm$^{-2}$)                   & $3.2 \times 10^{13}$ & $>5.5 \times 10^{13}$ & $3.0 \times 10^{13}$ & $>1.0 \times 10^{14}$ & $> 2.5 \times 10^{13}$ & $1.3 \times 10^{13}$ \\

$N(CO)~^b$   (cm$^{-2}$)                     & $8.0 \times 10^{15}$ & $1.5 \times 10^{17}$ & $1.3 \times 10^{16}$ & $4.6 \times 10^{16}$ & $2.8 \times 10^{17}$ & $1.6 \times 10^{16}$ \\

$N(o-H_2O)/N(H_2)~^c$                      & $< 2.0 \times 10^{-5}$  & $< 3.6 \times 10^{-7}$ & $< 3.3 \times 10^{-6}$ & $< 1.5 \times 10^{-6}$ & $< 8.2 \times 10^{-8}$ & $< 2.8 \times 10^{-6}$ \\

$N(o-H_2O)/N(H_2)~^d$                      & $4.0 \times 10^{-7}$  & $>3.6 \times 10^{-8}$ & $  2.3 \times 10^{-7}$ & $>2.2 \times 10^{-7}$ & $> 9.2 \times 10^{-9}$ & $8.1 \times 10^{-8}$ \\

\tableline\tableline \\
 
\end{tabular}}
$a$ Assumes $N$(H$_2^{16}$O)/$N$(H$_2^{18}$O) = 500.  \\ 
$b$ From LVG calculations (this paper) assuming N($^{12}$CO):N($^{13}$CO) = 55:1. \\
$c$ From the H$_2^{18}$O results assuming that CO:H$_2$ = 10$^{-4}$. \\
$d$ From the H$_2$O results assuming that CO:H$_2$ = 10$^{-4}$. \\
\end{table*}
}

%




%



\center{
\begin{table*}
\caption{Estimated OH Column Densities and Abundances in W49} 
\vskip 0.3 true in
 
\begin{tabular}{lcccc}
\tableline\tableline
   &      $V_{LSR}$ (\kms)      &       \\
   &      54.2         &  68.8  \\
\tableline

$\tau_o\Delta V(1665 MHz)$  (km s$^{-1}$) &   0.04 &  0.01 \\

$\tau_o\Delta V(1667 MHz)$  (km s$^{-1}$)    &  0.08 &  0.02 \\

$N(OH)$                       &  $1.4 \times 10^{14}$ &  $3.5 \times 10^{13}$ \\

$N(OH)/N(H_2)$     &  $1.1\times10^{-6}$ &  $2.3\times10^{-7}$ \\
 
$N(H_2O)_{tot}^a/N(OH)$  &   0.26  &   0.43 \\

\tableline\tableline \\

\end{tabular}
\\$^a$ - From Table 7, row 5 $\times 4/3$ to account for the ortho/para ratio.
\end{table*}
}

\clearpage 



\begin{figure*}[t]
\plotone{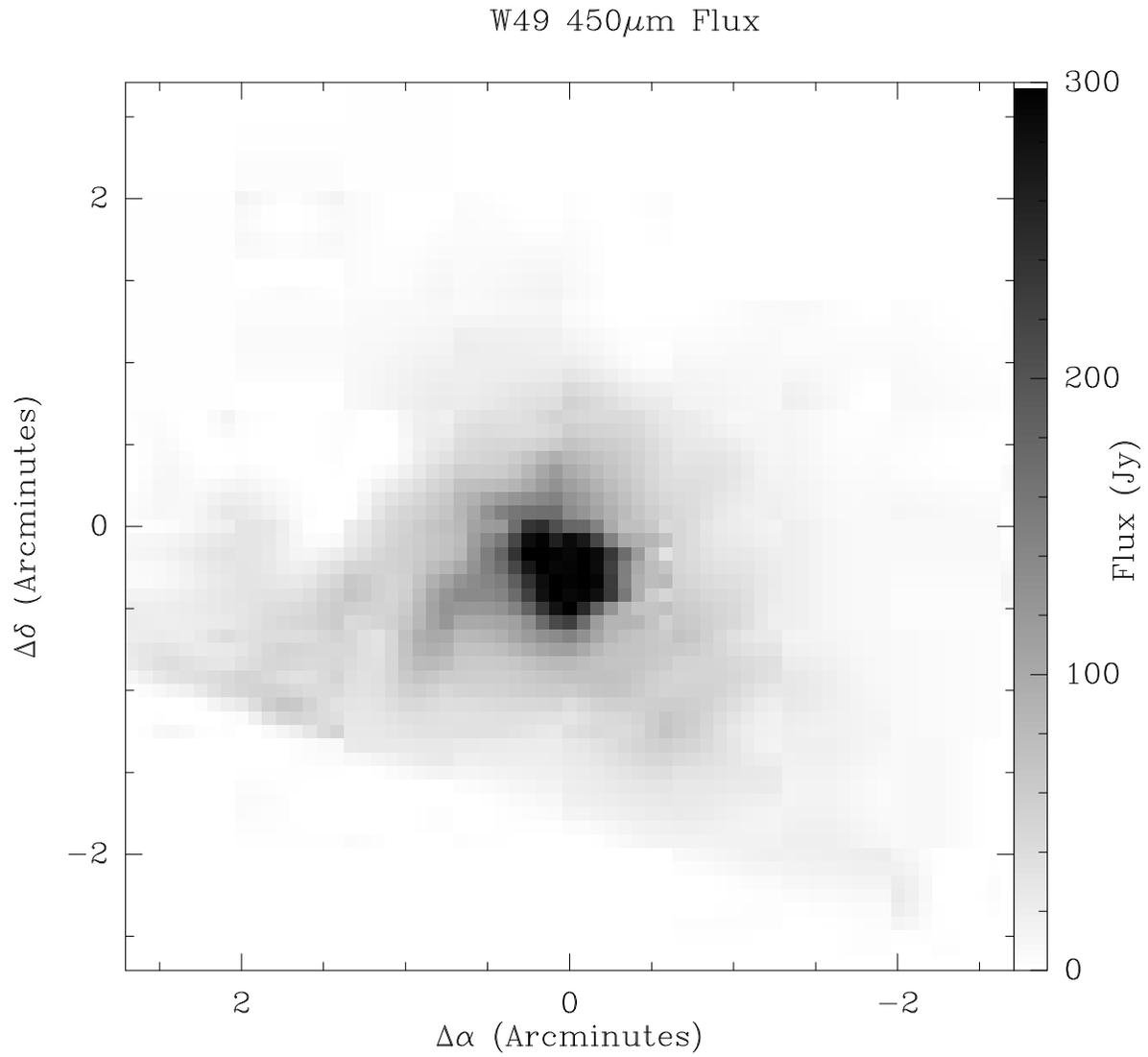} 
\vskip 0 in
\figcaption{450 $\mu$m continuum emission from W49A from SHARC at the CSO with a beamsize of approximately 9$''$.  The 
peak flux is $\sim$ 700 Jy but the image scale is truncated at 300 Jy to better show the weak, extended emission.
\label{fig:cont}}
\end{figure*}

\begin{figure*}[t]
\epsscale{.90}
\plotone{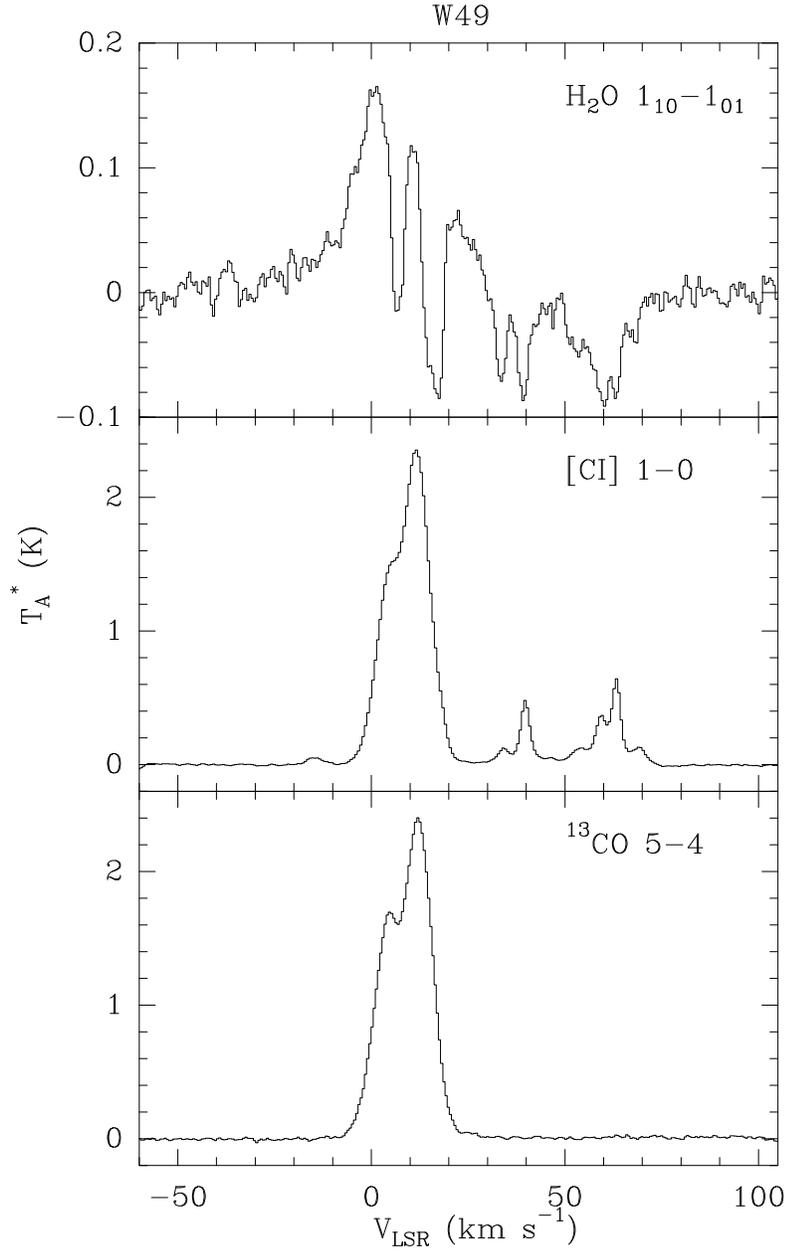} 
\vskip 0 in
\figcaption{Comparison of the 3 spectral lines detected by SWAS in W49A.
(top) H$_2^{16}$O $1_{10} - 1_{01}$. (middle) [C~I] $^3$P$_1 - ^3$P$_0$.
(bottom) $^{13}$CO ~$J = 5 \rightarrow 4$.  
\label{fig:spec1}}
\end{figure*}

\begin{figure*}[t]
\epsscale{0.8}
\plotone{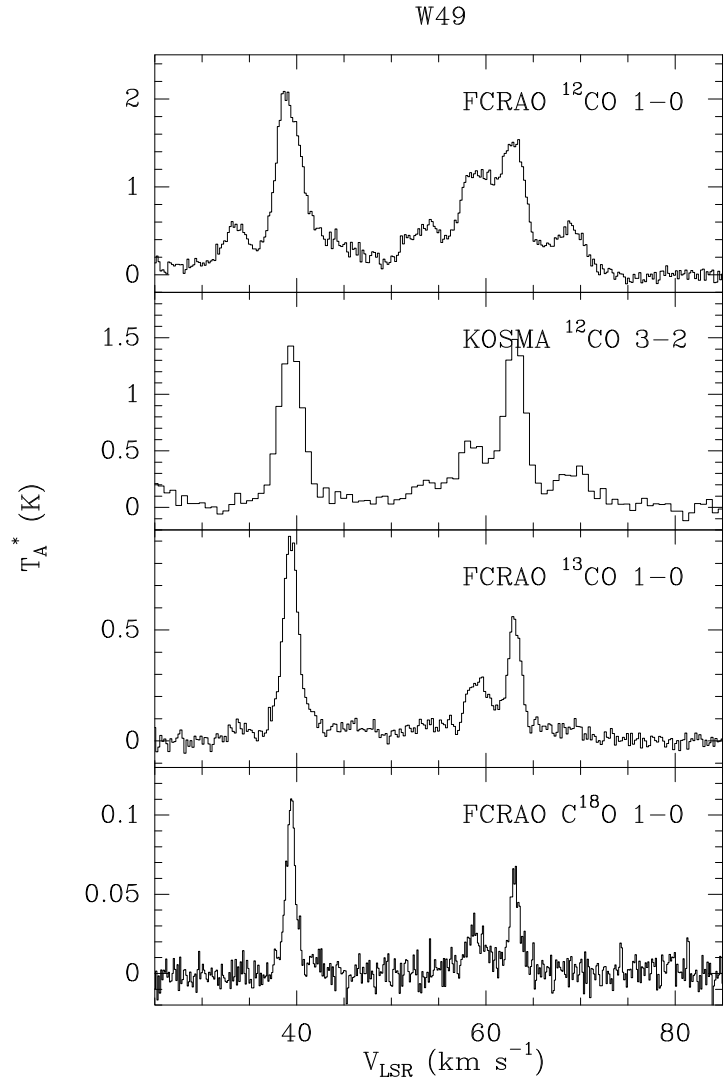} 
\vskip 0 in
\figcaption{Comparison of the FCRAO and KOSMA observations of W49A.  All
spectra have been smoothed to the angular resolution of the SWAS $^{13}$CO  $J = 5 \rightarrow 4$ line ($\sim 3.8'$).
\label{fig:spec2}}
\end{figure*}

\begin{figure*}[t]
\plotone{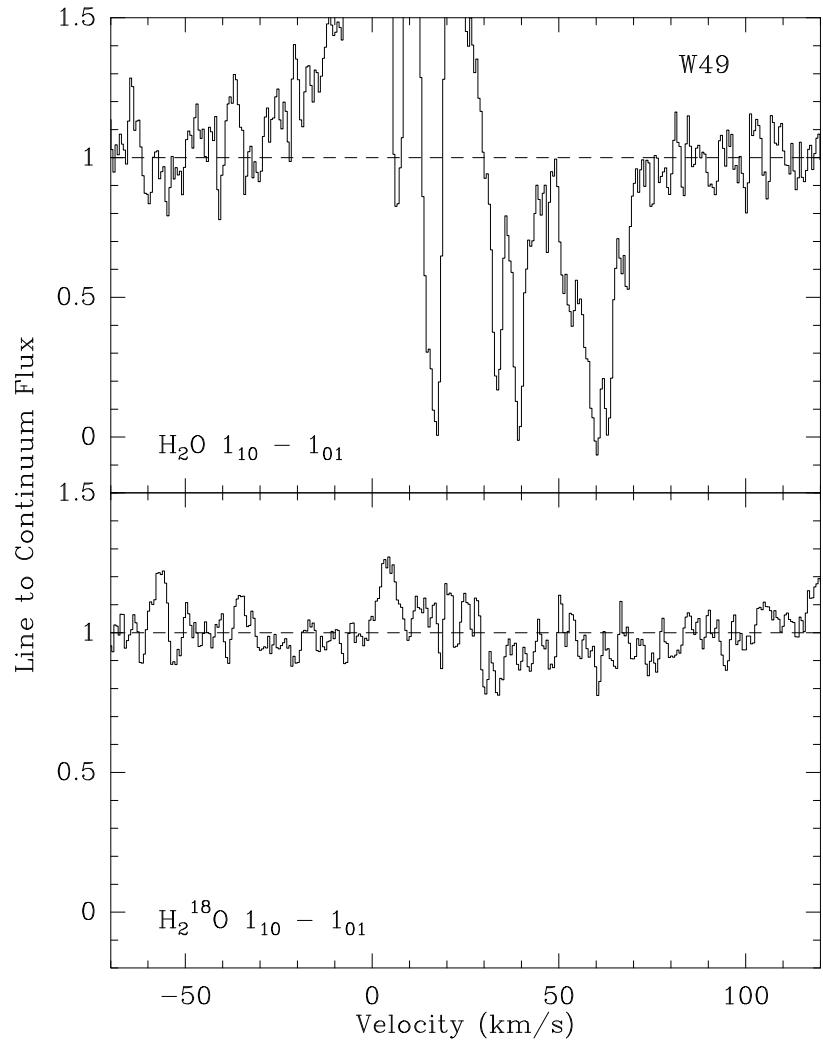} 
\vskip 0 in
\figcaption{
Line to continuum flux ratio in W49A for the SWAS observations of 
(top) H$_2^{16}$O $1_{10} - 1_{01}$ and (bottom) H$_2^{18}$O $1_{10} - 1_{01}$.
To account for slight tilts and curvatures in the raw spectra,  1st
 order baselines were subtracted from the  H$_2^{16}$O and
H$_2^{18}$O spectra, prior to calculating the line
to continuum flux ratios.
\label{fig:linecont}}
\end{figure*}

\begin{figure*}[t]
\plotone{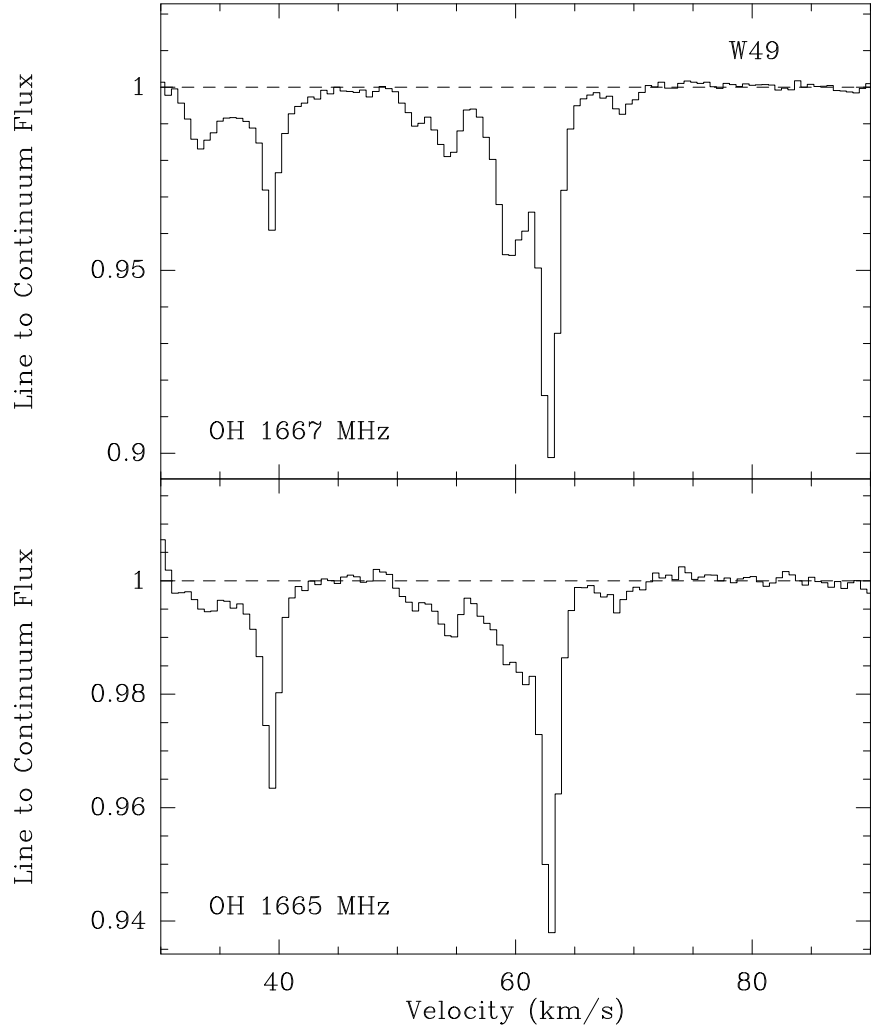} 
\vskip 0 in
\figcaption{
Line to continuum flux ratio in W49A for the Arecibo observations of the
(top) OH 1667 MHz and (bottom) OH 1665 MHz lines.
\label{fig:ohlinecont}}
\end{figure*}

\begin{figure*}[t]
\plotone{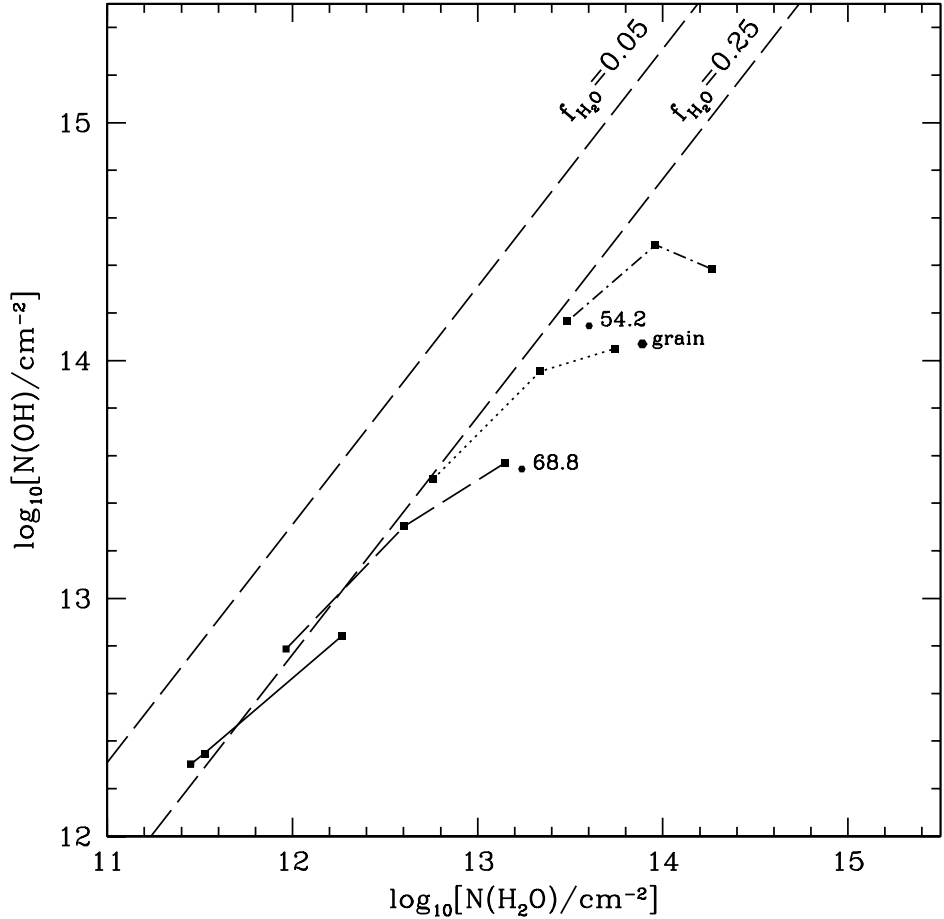} 
\vskip 0 in
\figcaption{
Comparison of observations/limits with results of PDR models. Observations are indicated by a black dot or  line
labeled with the OH velocity of the feature. Model results are shown for clouds with G$_o$ = 1.7 and  A$_V$=1 (solid curve), 2 (dashed curve),
3 (dotted curve), and 4 (dash-dotted curve). For each A$_V$, results are shown for cloud densities of 10$^2$, 10$^3$ and 10$^4$ cm$^{-3}$,
from left to right respectively along each curve. A single point (labeled as ``grain'') indicates the computed H$_2$O and OH column
densities from a model with $n=10^{2}$  cm$^{-3}$, $A_V=3$, and the catalytic formations of water on grain surfaces followed by photodesorption. Straight dashed lines follow from the analytical model of Neufeld et al. (2002), where the 
upper line is for  branching ratio $f_{\rm H_2O}:f_{\rm OH}:f_{\rm O}$ of 0.05:0.65:0.30 and the lower line is for 0.25:0.75:0.0. 
\label{fig:branch}}
\end{figure*}

\begin{figure*}[t]
\plotone{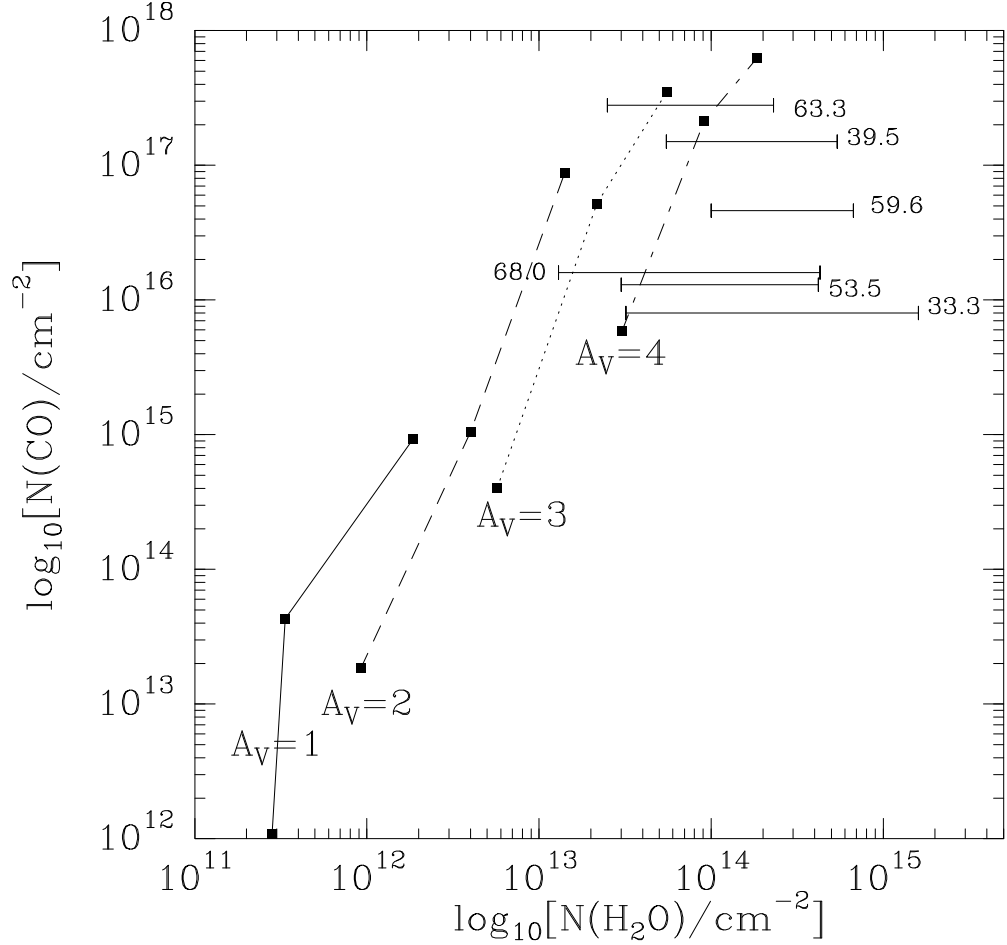} 
\vskip 0 in
\figcaption{
Comparison of the H$_2$O and CO observations with results of PDR models. Observations are indicated by the horizontal lines indicating the upper and lower limits of the H$_2$O column densities as determined from the H$_2^{18}$O and H$_2$O observations respectively, and are
labeled with the H$_2$O velocity of the feature. Model results are shown for clouds with G$_o$ = 1.7 and  A$_V$=1 (solid line), 2 (dashed line),
3 (dotted line), and 4 (dash-dotted line). For each A$_V$, results are shown for cloud densities of 10$^2$, 10$^3$ and 10$^4$ cm$^{-3}$,
from bottom to top respectively along each curve. 
\label{fig:h2oco}}
\end{figure*}

\begin{figure*}[t]
\plotone{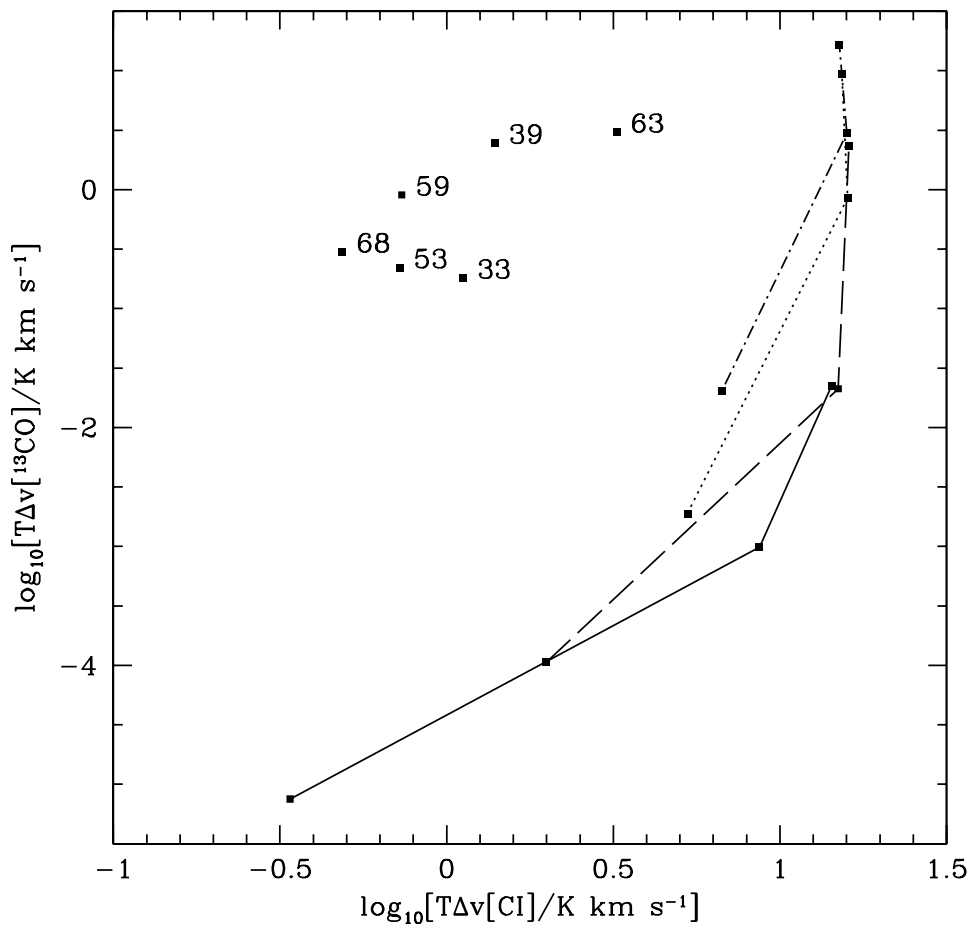} 
\vskip 0 in
\figcaption{
Comparison of observed [C~I] 492 GHz and $^{13}$CO $J = 1 \rightarrow 0$ emission with results of constant density PDR models. 
Curves show the model results, with the same line types as in the previous figure. Cloud densities are 10$^2$, 10$^3$ and 10$^4$ cm$^{-3}$ and increase from left to right.  Points indicate the observed [C~I] 
and $^{13}$CO integrated intensites, corrected as in the text, for the line-of-sight toward the 450$\mu$m continuum source (with a 1.5$'$ source size). 
\label{fig:cico}}
\end{figure*}
\end{document}